\documentstyle[12pt,epsf,cite]{article}
\begin{document}
\renewcommand{\thefootnote}{\fnsymbol{footnote}}
\begin{titlepage}
\begin{flushright}
KEK-TH-618 \\
March 1999 \\
\end{flushright}
\vspace*{10mm}
\huge
\begin{center}
Type IIB Random Superstrings 
\end{center}
\vspace{5mm}
\large
\begin{center}
Satsuki Oda \footnote[2]{Institution from April 1999 is KEK. E-mail address: oda@ccthmail.kek.jp}
and 
Tetsuyuki Yukawa {\small $^{\ddag,\S}$}
\footnote[0]{\hspace{-1.5mm}{\scriptsize $^{\ddag}$,$^{\S}$}
E-mail address: yukawa@koryuw02.soken.ac.jp,  yukawa@ccthmail.kek.jp}
\end{center}

\normalsize

\begin{center}
$^{\dag}$
Nara Women's University, Nara 630-8506, Japan
\vspace{0.5cm}

$^{\ddag}$
Coordination Center for Research and Education, \\
The Graduate University for Advanced Studies,   \\
Hayama-cho, Miura-gun, Kanagawa 240-0193, Japan 
\vspace{0.5cm}

$^{\S}$
High Energy Accelerator Research Organization (KEK),\\ 
Tsukuba, Ibaraki 305-0801, Japan
\end{center}
\vspace{0.5cm}
\begin{abstract}
\noindent
We consider random superstrings of type IIB 
in $d$-dimensional space.
The discretized action is constructed from the supersymetric matrix model,
which has been proposed as a constructive definition of superstring
theory.
Our action is invariant under the local $N=2$ super transformations,
and doesn't have any redundant fermionic degrees of freedom.
\end{abstract}
\end{titlepage}
  
Recent remarkable developments in superstring theory
are mostly due to duality and the D-brane~\cite{Duarity_D-brane_review}.
They shed light on the universal framework of the theory as well as on 
its non-perturbative face.
Following these discoveries, constructive definitions
of superstring were proposed~\cite{BFSS,IKKT},
which provide practical prescriptions of how to proceed 
beyond perturbative limits.
Those theories are expressed on the basis of matrices, and
are expected to restore the string picture in the limit of large matrix 
dimension, as it had been in the old matrix model~\cite{Matrix_model}.
Once we have a constructive definition of the theory,
it is a natural step to look for a proper regularization scheme 
in order to perform practical analysis.
As we have experienced in the course of developing QCD,
lattice regularization is the most powerful method for extending
the analysis into all regions of the coupling strength.
Understanding quantum gravity in terms of superstrings is commonly
recognized as being essentially non-perturbative.
In fact, several efforts to construct superstring theory 
on a lattice have been tried~\cite{Siegel_random_GS,Ambhorn_random_simulation}.
In those studies the Green-Schwarz superstring action 
was discretized while maintaining the local supersymmetry.
The problem in those descriptions is that the local $\kappa$-symmetry
is not explicitly fulfilled, and it is expected to recover
only in a suitable continuum limit.
In the recent constructive formulation of superstring based on 
the type IIB model,
the redundant fermionic degrees are removed by a special choice
of fermionic pairs in accordance with the local $\kappa$-symmetry.
The action of type IIB matrix model is then expressed 
in the semi-classical (i.e. large matrix size) limit by
\begin{equation}
S = \int d^2 \sigma \sqrt{g} \left[
    \frac{1}{4} \{ X^{\mu}, X^{\nu} \}^2
  - \frac{i}{2} \bar{\theta} \Gamma_{\mu} \{ X^{\mu}, \theta \} 
  + \lambda \right],
\end{equation}
where the Poisson brackets $\{ \;, \; \}$ are defined by
\begin{equation}
\{X,Y\} \equiv  \frac{1}{\sqrt{g}}\epsilon^{ab}\partial_a X \partial_b Y.
\end{equation}
Here, $X^{\mu}$ ($\mu = 1,2,\cdots,d$) are the $d$-dimensional 
space-time coordinates,
and $\theta^A$ ($A = 1, 2,\cdots$, {\it N}) 
are {\it N} sets of the anti-commuting spinor coordinates.
Also, $g$ is the absolute value of the determinant of world-sheet metric
($g_{ab}$), and $\epsilon^{ab}$ is an anti-symmetric tensor.
The $N=2$ supersymmetry manifests itself in $S$ under
\begin{equation}
\left\{
\begin{array}{lcl}
\delta^{(1)}X^{\mu}   & = & i \bar{\epsilon}_1 \Gamma^{\mu} \theta \\
\delta^{(1)}\theta \: & = & - \frac{1}{2} \{X^{\mu},X^{\nu}\}
                                          \Gamma_{\mu \nu} \epsilon_1,
\end{array}
\right.  \hspace{15mm}                                  
\left\{
\begin{array}{lcl}
\delta^{(2)}X^{\mu}   & = & 0      \\
\delta^{(2)}\theta \: & = &  \epsilon_2,
\end{array}
\right.  
                           \label{eq:Schild_con_SUSY_tr2}
\end{equation}
where $\Gamma^{\mu \nu}$ is an antisymmetric tensor defined by 
$\Gamma^{\mu \nu} = \frac{1}{2} [ \Gamma^{\mu}, \Gamma^{\nu} ] $.

In order to obtain the discretized action corresponding to $S$, 
we perform triangulations of world-sheet by 
equilateral triangles with side length $a$.
The action is a sum over contributions from each triangle
with three vertices $i$, $j$ and $k$:
$S = \sum_{\langle ijk \rangle} S^{\langle ijk \rangle}$.
On each triangle both bosonic and fermionic fields are approximated
by linear functions of local world-sheet coordinates 
$(\sigma^1,\sigma^2)$.
Since the Poisson brackets on each triangle are constant for linearly
approximated fields,
the integral over a triangle can be
trivially carried out and the discretized action is obtained as
\begin{equation}
S = S_B + S_F 
\sim \sum_{\langle ijk \rangle} \left( S_B^{\langle ijk \rangle} 
                                  + S_F^{\langle ijk \rangle} \right),
                                       \label{eq:Schild_dis_action}
\end{equation}
where
\begin{eqnarray}
S_B^{\langle ijk \rangle} 
 & = &   \frac{\triangle}{2} \cdot
         \frac{1}{4} \left( \frac{}{}
                        \{ X^{\mu}, X^{\nu} \}_{\langle ijk \rangle} 
                         \right)^2     \label{eq:dis_action_B_Poisson}    \\
 & = & \frac{1}{4 \triangle}  \left[ \;
     - \frac{1}{4} \{ (X_{ij}^2)^2 + (X_{jk}^2)^2 + (X_{ki}^2)^2 \} 
     + \frac{1}{2} \{ X_{ij}^2 X_{jk}^2 
       + X_{jk}^2 X_{ki}^2 + X_{ki}^2 X_{ij}^2 \}  \; \right], 
                        \nonumber    \label{eq:Schild_Dbos_action} \\
 &   & \nonumber \\
S_F^{\langle ijk \rangle} 
 & = & \frac{\triangle}{2} \left[ \; 
          -   \frac{i}{4} \bar{\theta}_{\langle ijk \rangle} \Gamma^{\mu} 
                             \{ X_{\mu}, \theta \}_{\langle ijk \rangle}
          -   \frac{i}{4} \{ \bar{\theta} , X_{\mu} \}_{\langle ijk \rangle} 
                            \Gamma^{\mu} \theta_{\langle ijk \rangle}
                                                  \;   \right] 
                                        \label{eq:dis_action_F_Poisson}  \\
 & = & \frac{1}{4 \triangle} \left[ \;
     - \frac{\triangle}{3} \{ \Omega_{ij} \! \cdot \! (X_{jk} \! - \! X_{ki})
                       + \Omega_{jk} \! \cdot \! (X_{ki} \! - \! X_{ij})
                       + \Omega_{ki} \! \cdot \! (X_{ij} \! - \! X_{jk}) \}
                                       \; \right],
                           \nonumber    \label{eq:Schild_Dfer_action} 
\end{eqnarray}
$\theta_{\langle ijk \rangle} = \frac{1}{3} (\theta_i + \theta_j + \theta_k)$
is the average strength of fermionic field for the triangle
$\langle ijk \rangle$,
and we denote 
$X_{ij}^{\mu} \equiv X_i^{\mu} - X_j^{\mu}$ and
$\Omega_{ij}^{\mu} \equiv \frac{i}{2} ( \bar{\theta}_i \Gamma^{\mu} \theta_j 
                      - \bar{\theta}_j \Gamma^{\mu} \theta_i )$.
Here, $X^{\mu}_i$ and $\theta_i$ are bosonic and fermionic fields 
on a vertex $i$, respectively,
and $\triangle$ is twice the area of elementary triangle,
i.e. $\triangle = \frac{\sqrt{3}}{2} a^2$.
The discretized Poisson brackets are written as
\begin{eqnarray}
\{A,B\}_{\langle ijk \rangle} 
& \equiv & \frac{1}{\triangle} f_{ijk} (A_{jk}B_{ki} - B_{jk}A_{ki})  \nonumber \\
& = & \frac{2}{\triangle} f_{ijk}
      ( A_{ij} B_{\langle ij \rangle} + A_{jk} B_{\langle jk \rangle} 
      + A_{ki} B_{\langle ki \rangle} ),                   
                                             \label{eq:Schild_dis_Poisson}
\end{eqnarray}
where
\begin{equation} 
f_{ijk}  = 
\left\{
\begin{array}{cl}
+1 & (i,\; j, \;k) \; {\rm counter \; clockwise} \\
-1 & (i,\; j, \;k) \; {\rm clockwise},             
\end{array}
\right. 
\end{equation}
and where the symmetric pair field is defined as
$A_{\langle ij \rangle} =  \frac{1}{2} (A_i + A_j)$.
The statistical property of the surface generated 
by the discretized bosonic action
was studied previously in \cite{Amb_Dur_Fre},
while the fermionic part was proposed in \cite{Siegel_random_GS} 
separately.
To show the $N=2$ supersymmetry, we needed to rewrite the transformation
so as to be valid in a discretized world-sheet.
Straightforward discretization of eq.~(\ref{eq:Schild_con_SUSY_tr2}) leads to
\begin{equation}
\left\{
\begin{array}{lcl}
\delta^{(1)}X_i^{\mu} & = & i \bar{\epsilon}_1 \Gamma^{\mu} \theta_i \\
\delta^{(1)}\theta_{\langle ijk \rangle} \: 
& = & - \frac{1}{2} \{ X^{\mu},X^{\nu} \}_{\langle ijk \rangle} 
                                   \Gamma_{\mu \nu} \epsilon_1, 
                                    \label{eq:Schild_dis_SUSY_tr1} 
\end{array} 
\right. \hspace*{5mm} \left\{
\begin{array}{lcl} 
\delta^{(2)}X_i^{\mu} & = & 0   \\ 
\delta^{(2)}\theta_i \: & = & \epsilon_2.   \label{eq:Schild_dis_SUSY_tr2}
\end{array}
\right.
\end{equation} 
We note that the Poisson bracket is constant on each triangle, and thus
the variation of the fermionic field $\delta^{(1)} \theta_{\langle ijk \rangle}$ 
is also defined on the triangle.
Let us first consider the invariance under $\delta^{(2)}$,
where $\delta^{(2)}S_B$ and $\delta^{(2)}S_F$ vanish trivially.
Next, the variation caused by $\epsilon_1$ 
are calculated as 
\begin{eqnarray}
\delta^{(1)} S_B 
& = & \frac{i}{2 \triangle} \sum_{\langle ijk \rangle} f_{ijk}^2 
      \left[ \frac{}{} \; 
      \bar{\epsilon}_1 \Gamma_{\mu} \theta_{jk} \left\{ 
      X^{\mu}_{jk} X_{ki}^2 - X^{\mu}_{ki} (X_{jk} \! \cdot \! X_{ki}) 
                                \right\} \right. \hspace*{3.5cm} \nonumber \\
& - & \left. \frac{}{} 
     \bar{\epsilon}_1 \Gamma_{\mu} \theta_{ki} \left\{ 
    X^{\mu}_{ki} X_{jk}^2 - X^{\mu}_{jk} (X_{jk} \! \cdot \! X_{ki}) 
                                      \right\} \right], 
                                  \label{eq:Schild_delta_1_S_B_dis} \\
\delta^{(1)} S_F
& = & - \frac{i}{2 \triangle} \sum_{\langle ijk \rangle} f_{ijk}^2  
    \left[ \frac{}{} \; 
    \bar{\epsilon}_1 \Gamma_{\mu} \theta_{jk} \left\{ 
    X^{\mu}_{jk} X_{ki}^2 - X^{\mu}_{ki} (X_{jk} \! \cdot \! X_{ki}) 
                                      \right\} \right. \nonumber \\
& - & \! \left. \frac{}{} 
    \bar{\epsilon}_1 \Gamma_{\mu} \theta_{ki} \left\{ 
    X^{\mu}_{ki} X_{jk}^2 - X^{\mu}_{jk} (X_{jk} \! \cdot \! X_{ki}) 
                                      \right\} \right] 
    \! - \! \frac{1}{12} \sum_{\langle ijk \rangle}
    \bar{\epsilon}_1 \Gamma_{\mu} 
    \theta_{[ \, i} \bar{\theta}_j \Gamma^{\mu} \theta_{k \, ]}.
                                     \label{eq:Schild_delta_1_S_F_dis}
\end{eqnarray}
For the first term in $\delta^{(1)}S_F$,
the relationships,
\begin{equation}
\sum_{\tiny \begin{array}{c}k = 0 \\
         {\rm closed \; loop} \; (i_0=i_n)\end{array}}^{n-1} 
    X_{i_k \, i_{k+1}}^{\mu} = 0, \hspace*{5mm}
3 \delta^{(1)} \theta_{\langle ijk \rangle}
  = 2 \delta^{(1)} \theta_{\langle ij \rangle} +  \delta^{(1)} \theta_k,
\end{equation}
guarantee ``integration by parts'' in the discretized form.
The first term of the $\delta^{(1)} S_F$
in eq.~(\ref{eq:Schild_delta_1_S_F_dis}) is canceled out 
by the $\delta^{(1)} S_B$ shown in eq.~(\ref{eq:Schild_delta_1_S_B_dis}).
Remainder of $\delta^{(1)} S$ is written as
\begin{equation}
\delta^{(1)} S
    = - \frac{1}{12} \sum_{\langle ijk \rangle}
    \bar{\epsilon}_1 \Gamma_{\mu} \theta_{[ \, i} 
    \bar{\theta}_j \Gamma^{\mu} \theta_{k \, ]}. \\
\end{equation}
This $\delta^{(1)} S$ 
vanishes in the cases of $D$ = 3, 4, 6 or 10
by making use of properties of $\Gamma$ matrices.
For the local supersymmetry, 
we must pay attention to the one-to-one correspondence between
the variations of the fermionic variables on
the triangles and on the vertices.
For this purpose we have checked whether
there exist $N_0$ independent relations among 
the $N_2 = 2 (N_0 - \chi)$ equations,
\begin{equation}
\delta^{(1)} \theta_{\langle ijk \rangle}= \frac{1}{3} \left(
\delta^{(1)} \theta_i + \delta^{(1)} \theta_j + \delta^{(1)} \theta_k \right),
\end{equation}
for sufficiently large surfaces with the Euler characteristic $\chi$.
In order to count how many independent relations exist among
the $N_2$ relations,
we compute numerically orthonormal vectors with $N_0$ components
by Schmidt's method. 
For most of the large $N_0$ configurations except those which have very
high spacial symmetry,
there exist $N_0$ orthonormal vectors,
and thus there mostly exists 
the one-to-one correspondence between triangles and vertices 
for sufficiently large surfaces.
For configurations with extra space symmetry, 
for example the discrete rotational symmetry, however,
the number of orthonormal vectors becomes
less than $N_0$.
Because these configurations are expected to have degeneracy,
we need to include the symmetry factor
in the integration measure, or extra factor to insure the detailed balance
in numerical simulations.

Quantization is carried out by the standard path-integral method.
The integration over $\sqrt{g}$ is interpreted as the sum over 
dynamical triangulations.
For the type IIB supersring,
the discretized partition function with a fixed number of vertices is given by
\begin{equation}
Z(N_0) = \sum_{\tiny {\rm triangulation}\:{\cal T}} 
         \frac{1}{\cal S(T)} \int \prod_{i=0}^{N_0-1}
         {\rm d} X_i {\rm d} \bar{\theta}_i {\rm d} \theta_i e^{- S_B - S_F} 
         \delta (X_0) \delta (\bar{\theta}_0) \delta (\theta_0),
\end{equation}
where ${\cal S(T)}$ is the symmetry factor of the triangulation (${\cal T}$)
to take into account the degeneracy.
It is noticed that the modes associated with the translational 
invariance of the action for bosonic fields and the $\delta^{(2)}$
supersymmetry for fermionic variables have been eliminated 
by fixing at zero for variables on $i=0$ vertex.
In the following we consider the case of a tetrahedron 
world-sheet, which corresponds to the smallest tree diagram.
In this case the triangulation is unique,
\begin{figure}
\epsfxsize=6cm \epsfysize=6cm
\centerline{\epsfbox{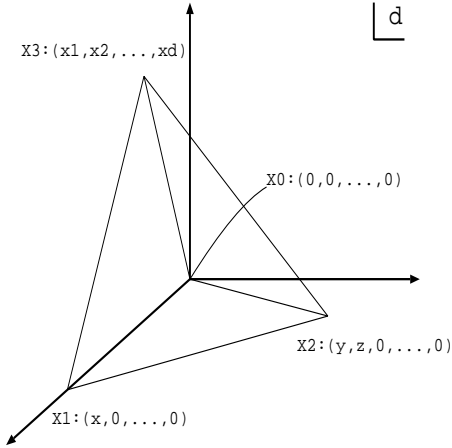}}
\caption
{
Tetrahedron.
}
\label{Fig:tetra}
\end{figure}
since a configuration with four vertices is only a tetrahedron
in our convention of not allowing tadpoles and self-energies.
By fixing the vertex $X_0$ at the origin and the triangle 
$\langle 012 \rangle$ on the $x$-$z$ plane
as shown in Fig.~\ref{Fig:tetra},
their integration measures are defined by 
\begin{eqnarray}
{\rm d} X_1 & = & x^{d-1} dx d \Omega_{d-1},   \nonumber \\
{\rm d} X_2 & = & dy z^{d-2} dz \Omega_{d-2},  \\
{\rm d} X_3 & = & \prod_{i=1}^d dx_i \; .     \nonumber
\end{eqnarray}
By integrating the fermionic coordinates 
out,\footnote{
We use the relation, 

$
\det \left( a^{\mu} b^{\nu} c^{\rho} \Gamma_{\mu \nu \rho} \right)
= \left[ - a^2 b^2 c^2 + (a \! \cdot \! b)^2 c^2 
  + (b \! \cdot \! c)^2 a^2 + (c \! \cdot \! a)^2 b^2 
  - 2 (a \! \cdot \! b) (b \! \cdot \! c) (c \! \cdot \! a) 
                                                      \right]^{\frac{k}{2}},
$
}
the partition function is given by
\begin{equation}
Z(4) 
 \propto \int_0^{\infty} x^{d-1} z^{d-2} dx dz \int_{-\infty}^{\infty} dy
      \prod_{i=1}^{3} dx_i
      \left[ \frac{}{}(x_3^2 + \cdots + x_d^2) x^2 z^2 \right]^{\frac{k}{2}}
      e^{- S_B},
\end{equation}
where
\begin{eqnarray}
S_B & \cong & x^2 z^2 + (z x_1 - y x_2)^2 + x_2^2 x^2 
        + \{ (y-x) x_2 - (x_1 - x) z \}^2       \nonumber \\
    & + & 2 (x_3^2 + \cdots + x_d^2) \{ x^2 + y^2 + z^2 -xy \},
\end{eqnarray}
and $\Gamma^{\mu}$ are $k \times k$ matrices.
Integrating over $x_i$ and $y$, and defining
$X^2         = x^2 z^2 $ and
$\tan \theta = \frac{2}{\sqrt{3}} \left| \: \frac{z}{x} \: \right|$,
\begin{equation}
Z(4) \propto  \int_0^{\infty} dX |X|^{\frac{1}{2} (d+k-2)}
            e^{- \frac{4}{3} X^2} \int_0^{\frac{\pi}{2}} d \theta 
            (\cos \theta)^{\frac{1}{2}(d+k-4)}
            (\sin \theta)^{\frac{1}{2}(d+k-6)}.
                               \label{eq:Schild_tet_Z_last}
\end{equation}
From eq. (\ref{eq:Schild_tet_Z_last}),
the partition function of the tetrahedron is found to be finite 
for $d+k \geq 5$.
Thus, it shows that the fermion suppress the divergence.
However, in this case,
even the partition function of the bosonic string
becomes finite when $d \geq 5$. 
We have checked numerically that fluctuation of the mean square 
distance from the center of mass diverges like the random
walk for $d \leq 4$ cases while it converges for $d \geq 5$ cases.

\begin{figure}
\epsfxsize=8.0cm \epsfysize=7cm
\centerline{\epsfbox{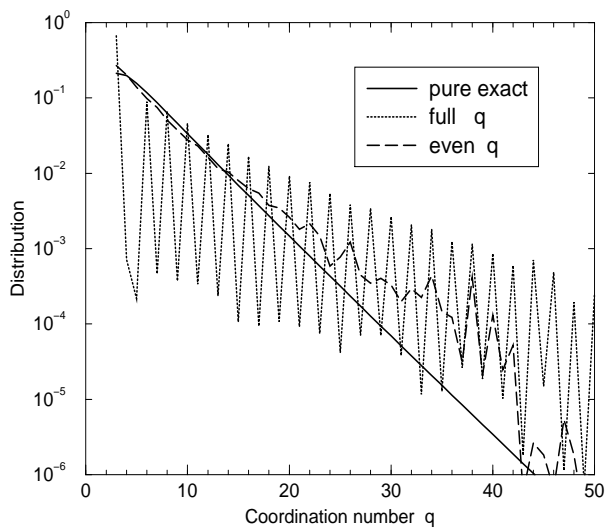}}
\caption
{
Dotted line and dashed line representing
the experimental coordination number distributions, 
$P_{N_0}(q)$ and $\tilde{P}_{N_0}(q)$, respectively,
for $d=5$ and $N_0=2000$ case.
The solid line denotes the theoretical solution 
for pure gravity~\protect \cite{Coord_dis_exact}.
}
\label{Fig:order_5s_40000}
\end{figure}

However, this is not so simple for larger size configurations.
In fact, we have found numerically 
that the partition function of the bosonic
string shows the spike singularity~\cite{Amb_Dur_Fre} 
for large $N_0$ in the case of $d=5$.
We have calculated the coordination number distributions $P_{N_0}(q)$
for the $d=5$ case which is shown by the dotted line 
in Fig.~\ref{Fig:order_5s_40000}.
We observe the zigzag structure where at the coordination number $q=3$ 
and at even $q$ values for $q \geq 6$ 
it shows peaks and at $q=4$ and at odd $q$ values for $q \geq 5$
it shows valleys.
When we pick up even coordination numbers and calculate
\begin{figure}
\epsfxsize=8cm \epsfysize=6cm
\centerline{\epsfbox{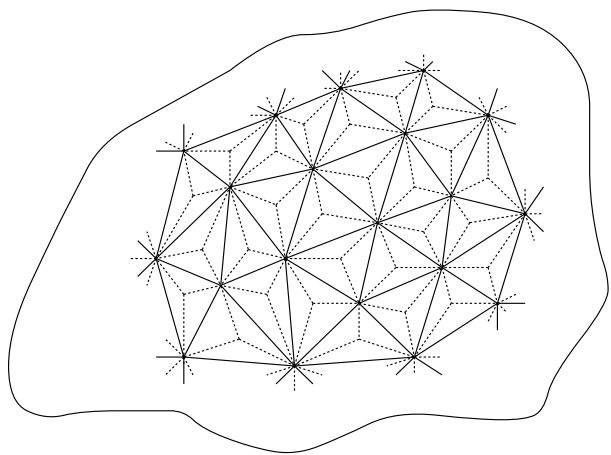}}
\caption
{
Typical spike configuration in terms of the world-sheet.
}
\label{Fig:Schild_bos}
\end{figure}
\begin{equation}
\tilde{P}_{N_0}(q) = 
       \frac{1}{\tilde{N_0}} \sum_i^{N_0} \delta_{2q,\: q_i} \hspace*{5mm} 
{\rm for} \;\; q \geq 3,
\end{equation}
where $\tilde{N_0} = \sum_{q=3}^{\infty} \sum_i^{N_0} \delta_{2q,\: q_i}$,
and $q_i$ denotes the coordination number of vertex $i$,
we obtain the distribution
shown by the dashed line in Fig.~\ref{Fig:order_5s_40000}.
It is similar to the distribution for pure gravity~\cite{Coord_dis_exact},
as shown by the solid line.
From this result, we can imagine that
the configurations of the bosonic string
are something like the one illustrated in Fig.~\ref{Fig:Schild_bos}.
The solid lines in Fig.~\ref{Fig:Schild_bos} are expected to describe 
almost like the same configuration as the pure gravity case.
If the fields on vertices around a vertex $i$ 
with coordination number $q_i=3$ shrink into a point,
its area becomes zero for any values of $X_i^{\mu}$.
Thus, we expect even for the case of $d > 5$, the partition function diverges
for a large $N_0$ system.
In order to avoid these divergences,
we should consider the superstrings,
because it is expected that these diverges are caused by tachyons of 
the bosonic fields,
and it will be suppressed by the contribution of fermionic fields 
of superstrings.

In conclusion, among several possible models of superstrings,
we have considered the type IIB superstrings.
We have proposed the discretized type IIB superstring action, 
and have shown that it is invariant 
under the local $N=2$ super transformations.
In the Green-Schwarz superstring action,
there contains a topological term for the local $\kappa$-symmetry
in order to separate the extra fermionic degree of freedom.
Existence of the topological term makes it difficult to discretize 
the Green-Schwarz superstring.
On the other hand, in the type IIB superstring which we have employed, 
the action has no extra fermionic degree,
which has been essential in the numerical study.
The partition function for the
tetrahedron configuration of the world-sheet
has shown
that effects of the fermionic fields indeed suppress the divergence,
and suggest that it also make the surface smooth for large systems.
For further studies, 
we need to find a powerful method 
which provide a method to calculate the fermionic determinant quickly
for a larger system.

\section*{Acknowledgements}
We wish to show special thanks to Prof.~H.~Kawai, N.~Ishibashi,
F.~Sugino and M.~Sakaguchi for many helpful discussions.
S.~Oda acknowledges Research Fellowships of JSPS

\end{document}